\documentclass[final,3p,times]{elsarticle}
\usepackage{xcolor}
\usepackage{amssymb}
\usepackage{natbib} 
\biboptions{square,sort&compress,comma}

\usepackage[LGRgreek]{mathastext}

\usepackage{lineno}
\usepackage{setspace}
\begin{document}

\begin{frontmatter}

\title{Investigation of magnetic ordering with spin reorientation transition and optical properties in Dy\textsubscript{2}CoCrO\textsubscript{6} nanomaterials}

\author{M. M. Islam \textit{$^{a}$},  M. A. Islam \textit{$^{b}$}, Rana Hossain \textit{$^{a,c}$}}
\author{M. J. Hosen \textit{$^{d}$}\corref{mycorrespondingauthor}}
\author{M. D. I. Bhuyan \textit{$^{a}$}\corref{mycorrespondingauthor}}
\address{
\textit{$^{a}$}Department of Physics, Mawlana Bhashani Science and Technology University, Santosh-1902, Bangladesh. \textit{$^{b}$Department of Physics, University of Chittagong, Chittagong, Bangladesh.}
\textit{$^{c}$Department of Mechanical Science and Bioengineering, Osaka University, Osaka 560-8531, Japan. \textit{$^{d}$Department of Physics, University of Dhaka, Dhaka-1000, Bangladesh.}}}


\cortext[mycorrespondingauthor]{Corresponding author}
\ead{jubaer@du.ac.bd, didar_bhuiyan@mbstu.ac.bd}


\begin{abstract}

This study reports the synthesis and physical properties of polycrystalline Dy$_{2}$CoCrO$_6$ (DCCO) nanoparticles. Analysis of the powder X-ray diffraction (XRD) pattern using Rietveld refinement showed that the compound crystallizes in an orthorhombic crystal structure with a space group \textit{Pbnm}. The particle size of approximately 57 nm was confirmed through micrographs obtained from field emission scanning electron microscopy and transmission electron microscopy. The X-ray photoelectron spectroscopy (XPS) investigations identified a mixed-valence state of Co and Cr cations. Magnetic susceptibility data indicated Curie-Weiss behavior in the temperature range 140-340 K and the onset of antiferromagnetic interactions with a Néel temperature of 119 K. At 31 K, DCCO shows spin reorientation transition from $\Gamma_4$(G$_x$A$_y$F$_z$) to $\Gamma_2$(F$_x$C$_y$G$_z$). Furthermore, room-temperature magnetization measurements demonstrated the antiferromagnetic behavior with weak ferromagnetic interactions at low temperatures. Additionally, DCCO exhibited semiconducting behavior with a direct optical bandgap of 1.97 eV, indicating promise for visible-light-driven energy harvesting and catalytic applications.

\end{abstract}

\begin{keyword}
\texttt{Perovskite oxides, Sol-gel method, Nanoparticles, Spin orientation, Semiconductor}
\end{keyword}

\end{frontmatter}

\section{Introduction}

Over the past few years, there has been a surge of research interest in mixed B site perovskite oxides, mainly attributed to their extensive range of physical, chemical, magnetic, and optical properties \cite{saha2020double,pomiro2016spin,sheikh2017lead,hossain2021oxygen}. With their diverse range of cationic orderings and oxidation states, mixed perovskite oxides, denoted as A$_2$BB$^{\prime}$O$_6$ or AB$_{0.5}$B$^{\prime}{_{0.5}}$O$_3$ (where A represents rare-earth elements and B/B$^{\prime}$ denote transition-metal elements), demonstrate a remarkable array of functional properties. These include superconductivity, a wide range of magnetic ordering, multiferroic behavior, and catalytic capabilities. As a result, mixed perovskite oxides hold immense potential for applications in multiple-state memory components, spintronics devices, switchable spin valves, high-frequency filters, magnetic field sensors, photovoltaic devices, photocatalytic systems, energy storage devices based on photosynthesis, and more \cite{vasala2015a2b,das2017pr2fecro6,yin2014multiferroicity,tanaka2001advances,bhalla2000perovskite}. Various mixed perovskite oxides were studied previously, both experimentally and theoretically, to explore their fascinating magnetic and optical properties \cite{pomiro2016spin,sheikh2017lead,yin2014multiferroicity,bhuyan2021sol,pan2018removal,hossain2021oxygen}.

Noticeably, rare-earth-based mixed perovskite oxides show complex magnetic structures such as half-metallicity, magnetic ordering, magnetoresistance, multiferroicity, magnetostrictive effect, etc., due to rich electronic structures and strong interaction between the 3d electrons along with strong spin-orbit coupling \cite{yin2014multiferroicity,bhuyan2021sol,kobayashi1998room,pomiro2016spin,hou2021insight}. For instance, an investigation on  YbFe$_{0.5}$Cr$_{0.5}$O$_3$ and TmFe$_{0.5}$Cr$_{0.5}$O$_3$ exhibits spin reorientation phenomenon due to transition of spin structure from $\Gamma_4$(G$_x$A$_y$F$_z$) to $\Gamma_2$(F$_x$C$_y$G$_z$) and a distinct magnetostrictive effect along with a negative thermal expansion that is ascribed to a magnetoelastic effect caused by repulsion between the magnetic moments of nearby transition metal ions \cite{pomiro2016spin}. Another investigation on DyFe$_{0.5}$Cr$_{0.5}$O$_3$ shows ferroelectric and antiferromagnetic ordering below transition temperature T$_{N1}$ = 261 K and strong magnetocaloric effect 11.3 J/kg.K at 4.5 tesla increased by magneto-electric interaction as a result of magnetic field and temperature-induced magnetic transition \cite{yin2014multiferroicity}. Recent investigation on RFe$_{0.5}$Cr$_{0.5}$O$_3$ (R = Nd and Sm) demonstrates an enormous exchange bias (EB) effect caused by the "pinning effect" of R on Fe/Cr spins as a result of the ferromagnetic (antiferromagnetic) interaction occurring between the R and Fe/Cr sublattices \cite{hou2021insight}.
Moreover, mixed perovskite oxide shows captivating optoelectronic properties, including a favorable bandgap, significant absorbance within the visible range of the solar energy, and great photocatalytic efficacy that makes them highly appealing \cite{sheikh2017lead,pan2018removal,hossain2021oxygen}. For instance, Ln$_2$NiMnO$_6$ (Ln = La, Eu, Dy, Lu) based solar cells show great photovoltaic performance due to a favorable bandgap very close to the Si (1.1 eV) \cite{sheikh2017lead}. Another investigation on La$_2$CoMnO$_6$ and La$_2$CuMnO$_6$ reveals great photocatalytic performance for the elimination of volatile organic compounds. For isopropyl alcohol, ethanol, toluene, and ethylene, 100\% conversion can be achieved with La$_2$CoMnO$_6$ at 150 $^{\circ}$C, 200 $^{\circ}$C, 300 $^{\circ}$C, and 350 $^{\circ}$C temperatures, respectively \cite{pan2018removal}. Recent investigation on DyFe$_{0.5}$Cr$_{0.5}$O$_3$ demonstrates bandgap narrowing due to two transition metals in B sites and the concurrent existence of oxygen deficiency \cite{hossain2021oxygen}. Such unique magnetic and optical properties motivated us to explore similar types of perovskite oxide Dy$_{2}$CoCrO$_6$ nanomaterials.
Nevertheless, it is challenging to produce B-site ordered perovskite oxides due to order-disordered effects during the synthesis governed by the thermodynamic and kinetic of the order-disorder reaction \cite{shimada2003kinetics,mandal2005new}. An investigation on R$_2$NiMnO$_6$ (R = La, Pr, Nd, Sm, Gd, Tb, Dy, Ho, and Y) demonstrates that perfect B site ordering can be achieved by optimizing the synthesis process \cite{booth2009investigation}. On the other hand, the order-disorder effect with various distortions in the perovskites framework is responsible for the complex physical behavior of mixed perovskite oxides \cite{bhuyan2021sol,feng2014high}. For instance, Gd$_2$FeCrO$_6$ perovskite oxide exhibits a complex state of antiferromagnetic (AFM) and ferromagnetic (FM) domain at a lower temperature due to octahedral distortion \cite{bhuyan2021sol}.
Furthermore, perovskite-type cobaltites have drawn significant attention because of the thermal fluctuation of their spin states \cite{knivzek2014non,dong2020giant}. The processes underlying the low-spin (LS, S = 0, t$^{6}_{2g}$e$^{0}_{g}$) to high-spin (HS, S = 2, t$^{4}_{2g}$e$^{2}_{g}$) transition of Co$^{3+}$ ion and its explanation regarding Co$^{3+}$ spin states have been debated for several decades \cite{dong2020giant}. Moreover, it is reliably confirmed that Co$^{3+}$ ion is in its non-magnetic low-spin state up to room temperature and a high temperature of 800 K, a simple LS-HS transition is observed \cite{knivzek2014non,dong2020giant}. 
Therefore, in the present investigation, we have prepared Dy$_{2}$CoCrO$_6$ for the first time using a cost-effective and straightforward sol-gel method to comprehensively explore their structural, morphological, magnetic, and optical characteristics extensively, along with potential applicability as a photocatalyst. 

\section{EXPERIMENTAL TECHNIQUES}
\subsection{Sample synthesis}
In this study, we employed a cost-effective sol-gel technique \cite{hench1990west,bokov2021nanomaterial} to synthesize DCCO perovskite oxide. Initially, Dy(NO$_3$)$_3$.6H$_2$O, Co(NO$_3$)$_2$.6H$_2$O, and Cr(NO$_3$)$_2$.6H$_2$O were taken in a stoichiometric ratio and dissolved individually in 40 ml of deionized water. Subsequently, the resulting clear solutions were combined. To form a polymeric-metal cation network \cite{chanda2015structural}, citric acid (CA) and ethylene glycol (EG) were added dropwise to this solution in a molar ratio of (Dy$^{3+}$, Co$^{3+}$+Cr$^{3+}$): (CA): (EG) = 1: 1: 4. To obtain the gel precursor, the solution was stirred at a temperature of 80 $^{\circ}$C for several hours. The gel was then completely burned at an elevated temperature, resulting in the formation of a powder sample. Finally, the obtained powder was ground and calcined at 800 $^{\circ}$C for 6 hours. The step-by-step synthesis process of this perovskite oxide is illustrated in the electronic supplementary information (ESI) figure S1.

\subsection{Characterizations} 
X-ray diffraction (XRD) data was collected by an XRD diffractometer (PW3040 X’pert PRO Philips) ranging from 10$^{\circ}$ to 80$^{\circ}$ with an X-ray source of Cu$_{K\alpha}$ = 1.5405 \AA. Then the analysis of XRD data was performed with the FullProf software package of Rietveld refinement. Fourier transform infrared spectroscopy (FTIR) was carried out to determine the functional group presence in the as-synthesized DCCO perovskite oxide using an FTIR spectrometer (PerkinElmer). Field-emission-electron-microscopy (FESEM) (JEOL JSM 7600F) and transmission-electron-microscopy (TEM) (Talos F200X) were performed to observe the surface morphology of as-produced DCCO perovskite oxide. An X-ray spectroscope was used to conduct an energy-dispersive X-ray (EDX) spectrum to observe the elemental composition of as-produced DCCO. To determine the valence state of the different elements that exist in DCCO perovskite oxides, X-ray photoelectron spectroscopy (XPS) was carried out by an Al-K$_\alpha$ (1486.6 eV) X-ray source. Temperature-dependent magnetization (M-T) measurement was investigated in both field-cooled (FC) and zero-field-cooled (ZFC) modes using a quantum design physical-properties-measurement-system (PPMS) DynaCool. Field-dependent magnetization was also observed using the same PPMS at temperatures 20, 100, 200, and 300 K. Absorbance spectrum of DCCO was determined by performing an ultraviolet-visible (UV-visible) spectrometer (UV-2600, Shimadzu). Finally, a photoluminescence (PL) spectrometer (RF-6000, Shimadzu) was used to verify the optical bandgap obtained from the absorbance spectrum. 

\begin{table*}
 \centering
\caption{\label{tab:table3} Structural parameters and reliability (R) factors of Dy$_{2}$CoCrO$_6$ perovskite oxide determined by Rietveld refinement.}
\begin{tabular}{c c c c c c c c}
\hline
Atoms & Wyc. positions & $x$ & $y$ & $z$& Bond length (\AA) & Bond angle ($^{\circ}$) &  R factors\\ \hline \\
 Dy&4c&$-0.01580$&$0.06246$&$0.25000$& Co/Cr-O1$=1.99$
 &$ <$ Co/Cr-O$1$-Co/Cr $>$&R$_{p}=1.98$ \\
 &&&&&&=$149.4$&\\
 Co&4b&$0.5$&$0$&$0$&Co/Cr-O2=$1.96$
 &$<$Co/Cr-O$2$-Co/Cr$>$&$R_{wp}=2.49$\\
 &&&&&&=$145.6$&\\
 Cr&4b&$0.5$&$0$&$0$&
 &&$\chi^2=1.35$\\
 O1&4c&$0.29418$&$ 0.29710$&$ 0.05145$&\\
 
 O2&8b&$0.10364$&$0.46210$&$0.25000$&\\
 \hline
 
\end{tabular}

\end{table*}
\begin{figure*}[h]
 \centering
 \includegraphics[width= 0.7\textwidth]{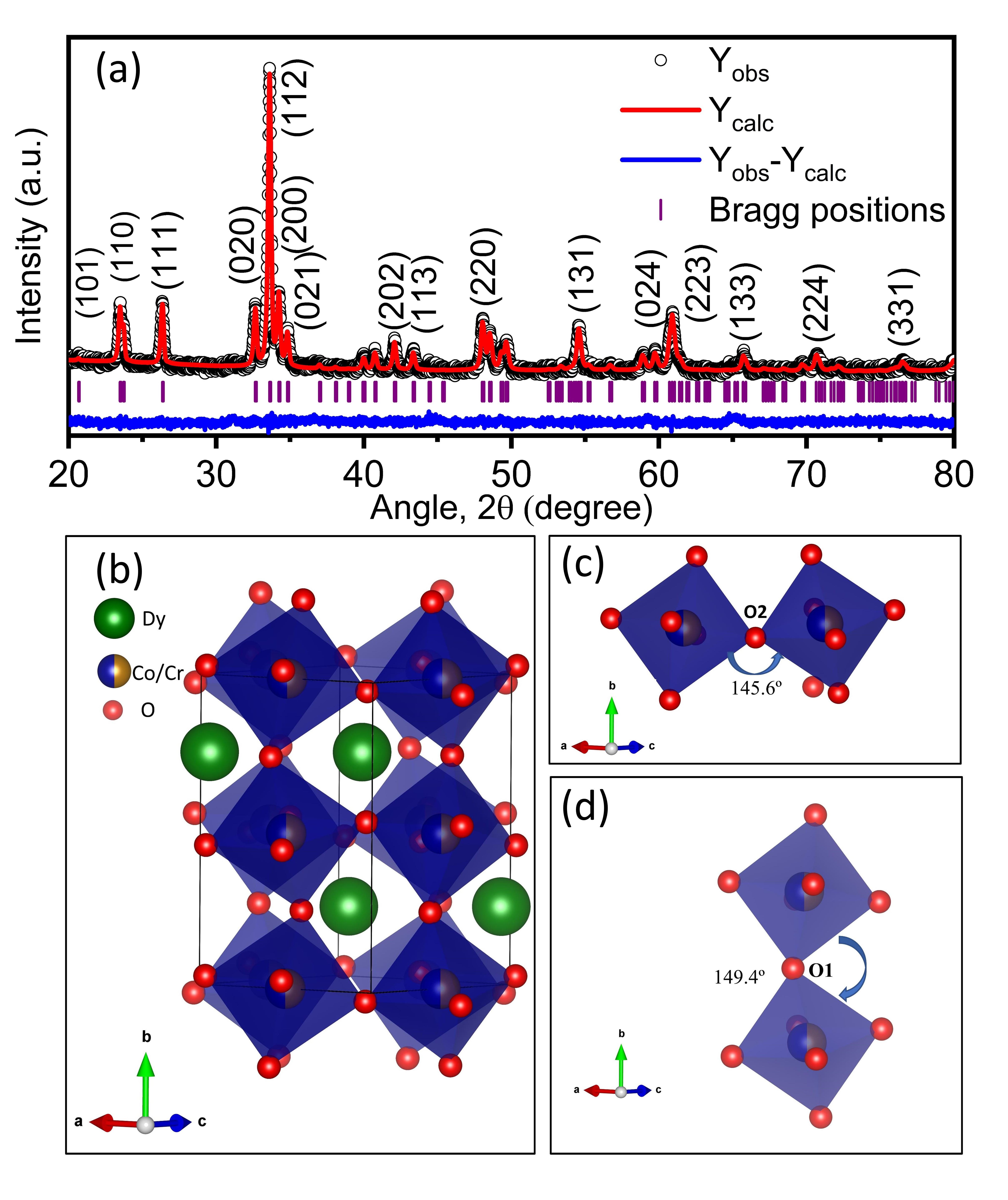}
 \caption{(a) Rietveld refined XRD spectrum of Dy$_{2}$CoCrO$_6$ perovskite oxide. (b) Unit cell and magnified image of linked Co/Cr-O octahedra (c and d).}\label{fig:xrd}
\end{figure*}


\begin{figure*}[h]
 \centering
 \includegraphics[width=0.8\textwidth]{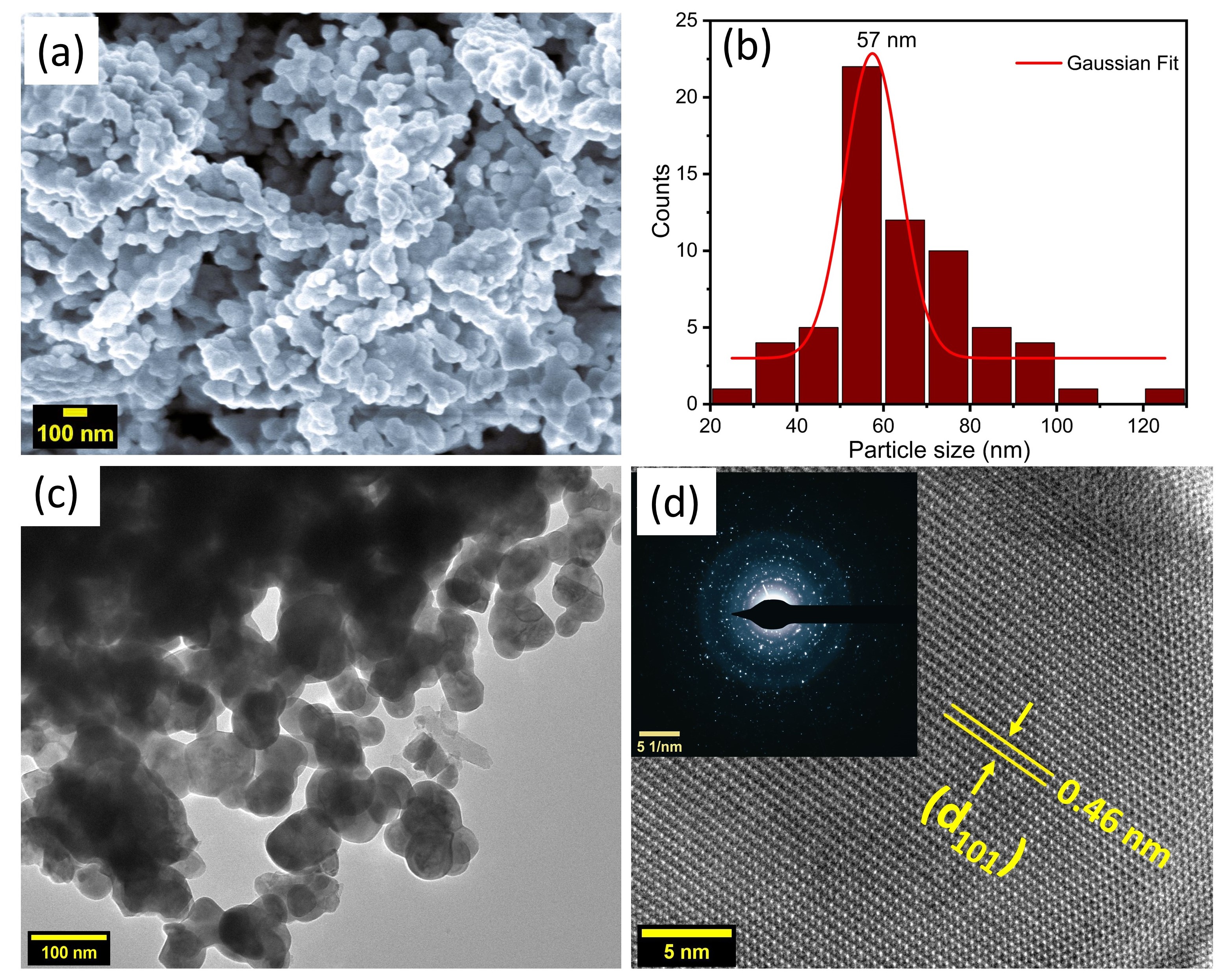}
 \caption{(a) \& (b) FESEM image and particles size distribution obtained from FESEM image, (b) \& (d) Bright field and High-resolution TEM images with SAED pattern in the inset of figure (d) of Dy$_{2}$CoCrO$_6$ nanoparticles.}\label{fig:morphology}
\end{figure*}

\section{RESULTS AND DISCUSSIONS}
\subsection{Crystallographic structure analysis}
To demonstrate the stability of as-synthesized DCCO perovskite material, we calculate the tolerance factor $t$ and octahedral factor $\mu$. The tolerance factor and octahedral factor can be defined by the following expressions \cite{jiang2021high}, 
\begin{eqnarray}
 t=\frac{R_{Dy}+R_O}{\sqrt{2(\frac{R_{Co}+R_{Cr}}{2}+R_O)}}
\end{eqnarray}
and 
\begin{eqnarray}
\mu =\frac{\frac{R_{Co}+R_{Cr}}{2}}{R_{O}} 
\end{eqnarray}
Here, R$_{Dy}$, R$_{Co}$, R$_{Cr}$, and R$_O$ represent the Shannon radii of Dy, Co, Cr, and O ions, respectively. Investigations on existing perovskite oxide \cite{jiang2021high,li2004formability} exhibit that the perovskite oxide can be formed for the tolerance factor and octahedral factor range 0.71 $< t <$ 1.0 and 0.42 $ < \mu < $ 0.75, respectively. In our investigation, the tolerance factor and octahedral factor of DCCO were found to be 0.92 and 0.57, respectively, which confirms that the perovskite structure of DCCO has been formed. Moreover, if $t$ is less than 0.97, the structure will be either orthorhombic or monoclinic \cite{anderson1993b}. Since the tolerance factor of DCCO perovskite oxide at room temperature (RT) was found to be 0.92, it is possible to predict its crystal structure as either orthorhombic or monoclinic.
Further, we have estimated the structural stability of DCCO perovskite oxide by calculating its global-instability-index (GII), which is described as the root-mean-square (rms) of the bond discrepancy factor calculated from bond valence sum \cite{lufaso2001prediction,yamada2018complementary},

\begin{eqnarray}
 GII=\sqrt{\frac{\sum_{i=1}^{N}(d_i)^2}{N}}
\end{eqnarray}

where, d$_i$ = the bond discrepancy factor and $N$ = ions number. Typically, for stable structures, the values of GII are lower than 0.1 valence unit (v. u.), and for unstable structures, the values of GII are greater than 0.2 v.u. \cite{lufaso2001prediction}. Especially, several investigations demonstrate that the stable perovskite oxides can be found under normal pressure for GII $<$ 0.02 v. u. \cite{li2018new,byeon2003high}. Noticeably, the GII value of DCCO was calculated to be 0.01 v. u., which is lower than 0.02 v. u., then it is feasible to form a stable structure of DCCO under atmospheric pressure.
For structural analysis, the Rietveld refinement of XRD patterns of DCCO was performed using the FullProf software package \cite{rodriguez1990fullprof}. No undesirable secondary peak was found in the XRD pattern, confirming the sample's high phase purity. Since the tolerance factor, $t$ is less than 0.97; hence we have considered both orthorhombic \textit{Pbnm} and monoclinic \textit{P2$_1$/n} space group for Rietveld refinement and found that the orthorhombic  
\textit{Pbnm} space group matches more accurately than the monoclinic \textit{P2$_1$/n} space group. In order to simulate the background, a series of background points with adjustable heights were interpolated linearly to create the background. By utilizing the pseudo-Voigt axial dispersion asymmetry of Thompson-Cox-Hastings, Bragg's reflections were modeled. Scale factor and cell parameters were refined following the refinement of full-width at half maximum (FWHM) and profile parameters. The thermal and spatial coordinates were refined after attaining correct profile matching \cite{bhuyan2021sol}. 
Figure ~\ref{fig:xrd} (a) shows the XRD spectrum of DCCO perovskite oxide attained from Rietveld refinement.

It is challenging to determine whether DCCO is ordered or not using the XRD technique solely because Co and Cr almost have the same ionic radii. However, one can learn about cation ordering by observing the superstructure peak at an angle of around 2$\theta$=20$^{\circ}$ \cite{shi2011local,chanda2016magnetic}. The DCCO superstructure peak at 20.2$^{\circ}$ is negligible, as seen in figure ~\ref{fig:xrd} (a), suggesting imperfect ordering at B-site. The mixed states of Co and Cr revealed by the XPS spectrum, which will be described later, also provide evidence of a lack of perfect ordering.

According to the Rietveld refinement, the DCCO perovskite oxide's lattice parameters are a = 5.231(1) \AA, b = 5.479(2) \AA, c = 7.496(3) \AA, with a cell volume of 214.87 \AA$^3$. We have used these lattice parameters to model the DCCO unit cell using the Visualization for Electronic and Structural Analysis (VESTA) software \cite{momma2011vesta,brown1985bond}. Figure ~\ref{fig:xrd}(b) depicts the DCCO unit cell, and figure ~\ref{fig:xrd}(c) and (d) depicts the magnified view of the Co/Cr-O octahedral obtained from VESTA. It is seen from Co/Cr-O octahedral that the Co and Cr ions are not at the octahedral center precisely, which indicates octahedral distortion. We have also computed tilt angle $\phi$ = $(180-\theta)/2$, where $\theta$ is the Co/Cr‒O‒Co/Cr bond angle to estimate the octahedral distortion \cite{das2017pr2fecro6}. The mean value of tilt angle $\phi$ is calculated to be 16.25$^{\circ}$, further confirming the structural distortion of the DCCO unit cell. Table 1 represents the structural parameters of DCCO perovskite oxide with the reliability (R) factors obtained from the Rietveld refined XRD pattern.

Moreover, FTIR analysis was conducted at RT to examine the chemical compound of as-prepared DCCO perovskite oxide, as shown in the ESI figure S2. Noticeably, the Co/Cr-O octahedral complex in the DCCO unit cell is anticipated to have six typical modes of vibration (V$_1$ - V$_6$). Among them, V$_6$ would be inactive, V$_3$ and V$_4$ would be IR active, and V$_1$, V$_2$, and V$_5$ would be Raman active \cite{nakamoto2009infrared,gaikwad2019structural}. The transition mode at about 516 cm$^{-1}$ in the FTIR spectrum of DCCO is related to V$_3$, which denotes the presence of bending vibrations of the Co/Cr-O bonds within Co/Cr-O octahedral. The transition band, V$_4$, near 575 cm$^{-1}$, is associated with the stretching vibrations of the Co/Cr-O bonds \cite{bhuyan2021sol}. The IR band around 775 cm$^{-1}$ is related to the trapped NO$^{3-}$ ions in the DCCO perovskite oxide \cite{nakamoto2009infrared}. The weak bands in the range of 1500–1700 cm$^{-1}$ and 3000–3600 cm$^{-1}$ exist due to the vibrations of C-O bonds and O-H bonds, which arise owing to the chemisorption of atmospheric CO$_2$ and adsorption of water by the surface of DCCO \cite{bhuyan2021sol,gaikwad2019structural}.

\subsection{Morphological and elemental analyses}
To investigate the surface morphology of the Dy$_{2}$CoCrO$_6$ perovskite oxide, FESEM, and TEM imaging were carried out of the sample. Figure ~\ref{fig:morphology}(a) shows the FESEM image of the Dy$_{2}$CoCrO$_6$ nanoparticles, and ~\ref{fig:morphology}(b) represents the distribution of the particles in terms of their corresponding size. From the FESEM imaging, the average particle size of DCCO perovskite oxide is approximately 57 nm. Furthermore, we have determined the average crystalline size using the Scherrer formula \cite{bhuyan2021sol} of these synthesized nanoparticles, and it is found to be around 30 nm. This size is considerably smaller than the particle size observed in the FESEM micrographs, indicating the possible occurrence of particle agglomeration. A bright field TEM image of the DCCO perovskite oxide is illustrated in figure ~\ref{fig:morphology}(c). The range of particle size 40–100 nm obtained from the TEM image is quite comparable with the data from the FESEM imaging. Additionally, the crystal lattice planes of DCCO nanoparticles with an interplanar separation of 0.46 nm, related to its (1 0 1) plane, are obtained from the high-resolution TEM (HRTEM) image displayed in figure ~\ref{fig:morphology}(d). In the inset of figure ~\ref{fig:morphology}(d), the ring-like pattern of selected-area-electron-diffraction (SAED) illustrates the polycrystallinity of DCCO perovskite oxide \cite{hosen2023structural}. 

\begin{table*}
\centering
\caption{Mass and atom percentages of Dy$_{2}$CoCrO$_6$ nanomaterials as attained by EDX spectrum.}
\begin{tabular}{lllll}
\hline
Elements & \begin{tabular}[c]{@{}l@{}}Mass (\%)\\ (Theoretical)\end{tabular} & \begin{tabular}[c]{@{}l@{}}Mass (\%)\\ (Experimental)\end{tabular} & \begin{tabular}[c]{@{}l@{}}Atom (\%)\\ (Theoretical)\end{tabular} & \begin{tabular}[c]{@{}l@{}}Atom (\%)\\ (Experimental)\end{tabular} \\ \hline 
Dy      & 61.10                                                             & 58.40                                                              & 20                                                                & 19.19                                                              \\
Co      & 11.08                                                             & 14.68                                                               & 10                                                                & 13.30                                                             \\
Cr      & 9.77                                                             & 9.67                                                               & 10                                                                & 9.93                                                               \\
O       & 18.05                                                             & 17.25                                                              & 60                                                                & 57.58                                                             \\ 
Total   & 100                                                               & 100                                                                & 100                                                               & 100  \\
\hline
\end{tabular}

\end{table*}

\begin{figure*}[h]
 \centering
 \includegraphics[width=0.7\textwidth]{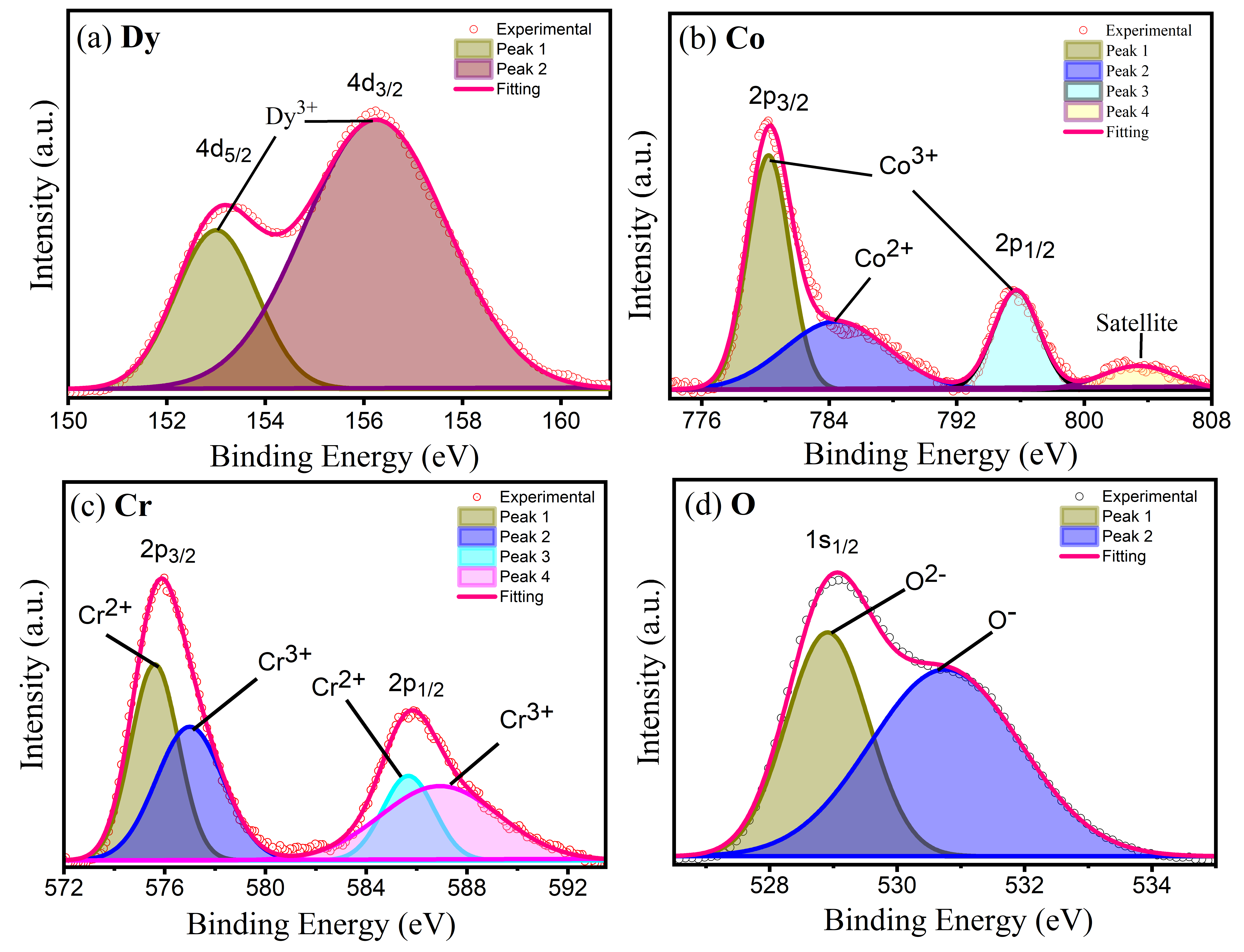}
 \caption{XPS spectra of (a) Dy, (b) Co, (c) Cr and (d) O elements of Dy$_{2}$CoCrO$_6$ nanoparticles.}\label{fig:XPS}
\end{figure*}

Moreover, we have analyzed the elemental constituent of as-prepared DCCO nanomaterials using the EDX spectrum at RT. ESI figure S3 demonstrates the EDX spectrum of as-prepared DCCO nanomaterials with the existence of all required elements. Table 2 shows the mass and atom percentage of Dy, Co, Cr, and O atoms present in the DCCO nanomaterials. The mass and atom percentages we obtained in our experiments closely aligned with the theoretical values, providing additional evidence of the successful synthesis of DCCO nanoparticles.

\subsection{XPS analysis}
X-ray photoelectron spectroscope (XPS) was used to examine the chemical binding energies and oxidation states of each atom of the as-prepared DCCO perovskite oxide. Figures~\ref{fig:XPS}(a-d) represent the core level XPS spectra of the Dy-4d, Co-2p, Cr-2p, and O-1s of the synthesized sample. The XPS spectrum of Dy 4d, as shown in figure ~\ref{fig:XPS}(a), exhibits two doublet peaks at 152.9 eV and 156.25 eV due to spin-orbit splitting, related to 4d$_{5/2}$ and 4d$_{3/2}$, respectively, confirming the existence of Dy$^{3+}$ state in the DCCO perovskite oxide \cite{michel2019novel}. The Co 2p orbital's spin-orbital splitting can be seen in the XPS spectrum of the Co content (Figure~\ref{fig:XPS}(b)) as two primary peaks near 780.1 eV and 795.85 eV, which correspond to the Co-2p$_{3/2}$ and Co-2p$_{1/2}$ states, respectively, and a satellite-like peak with the binding energy of 803.61 eV. Co-2p$_{1/2}$'s deconvolution peak at 795.85 eV represents the Co$^{3+}$ state, while Co-2p$_{3/2}$'s deconvolution peak at 780.1 eV and 784.5 eV represents the Co$^{3+}$ and Co$^{2+}$ states, respectively \cite{muscas2022nanostructure, ha20163dom}. In Dy$_{2}$CoCrO$_6$, the concentration of Co$^{3+}$ is 67\%, while the concentration of Co$^{2+}$ is 33\%. Figure ~\ref{fig:XPS}(c) shows the deconvolution of Cr 2p$_{3/2}$ at 575.5 eV and 577 eV, corresponding to Cr$^{2+}$ and Cr$^{3+}$ states and Cr 2p$_{1/2}$ at 585.58 eV and 586.5 eV, corresponding to Cr$^{2+}$ and Cr$^{3+}$ states, respectively \cite{maiti2013large, sarkar2016large}. By comparing the areas of the XPS spectra for the two valence states, [Cr$^{2+}$]/[Cr$^{3+}$] has been determined to be 0.76. Finally, figure ~\ref{fig:XPS}(d) demonstrates the XPS spectrum of the O-1s$_{1/2}$ orbital, the deconvolution peak near 529.02 eV is the characteristic peak of O$^{2-}$ ions, and the deconvolution peak near 530.58 eV describes the ionization of oxygen ions or O$^-$ species \cite{das2017pr2fecro6,hosen2023structural}.

\begin{figure*}[h]
 \centering
 \includegraphics[width=1.0\textwidth]{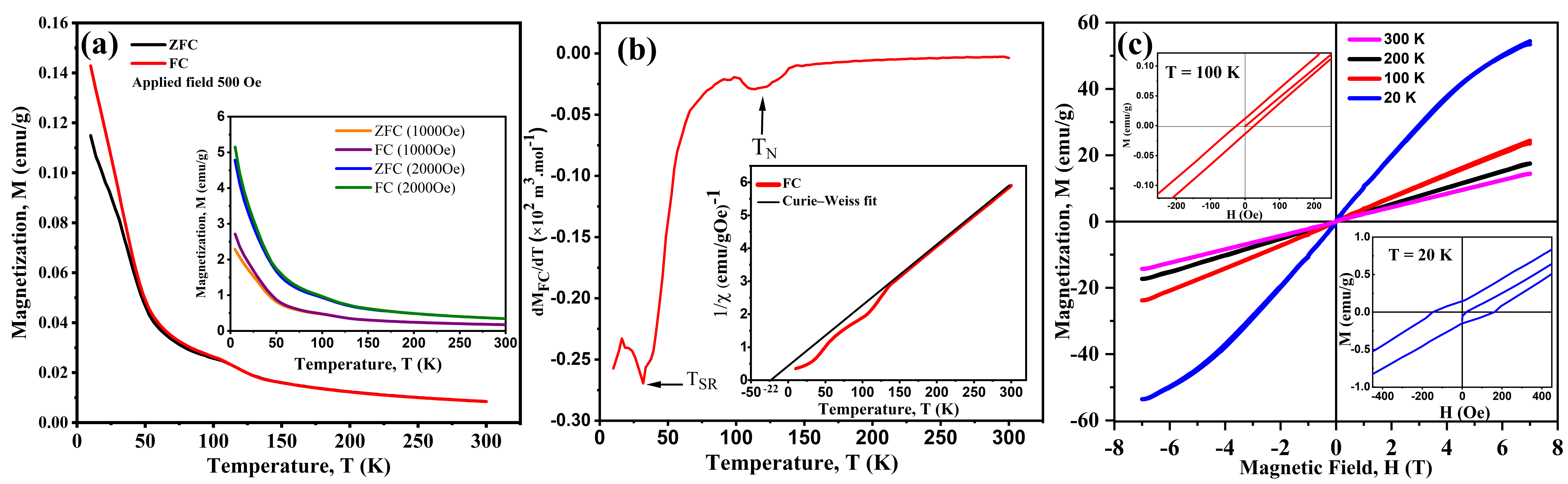}
 \caption{(a) FC and ZFC magnetization as a function of temperature under an external magnetic field of 500 Oe, 1000 Oe (inset) and 2000 Oe (inset). (b) The first derivative of the FC curve. Inset shows thermal dependence of inverse susceptibility of the compounds at H = 500 Oe, and (c) M-H curves determined at 300 K, 200 K, 100 K, and 20 K with a magnetic field applied up to $\pm$ 7 T. The upper and lower insets
show the magnified view of the hysteresis loops observed at 100 K and 20 K respectively.}\label{fig:Magnetic}
\end{figure*}

\subsection{Magnetic characterization}
\begin{figure}[h]
 \centering
 \includegraphics[width=0.7\textwidth]{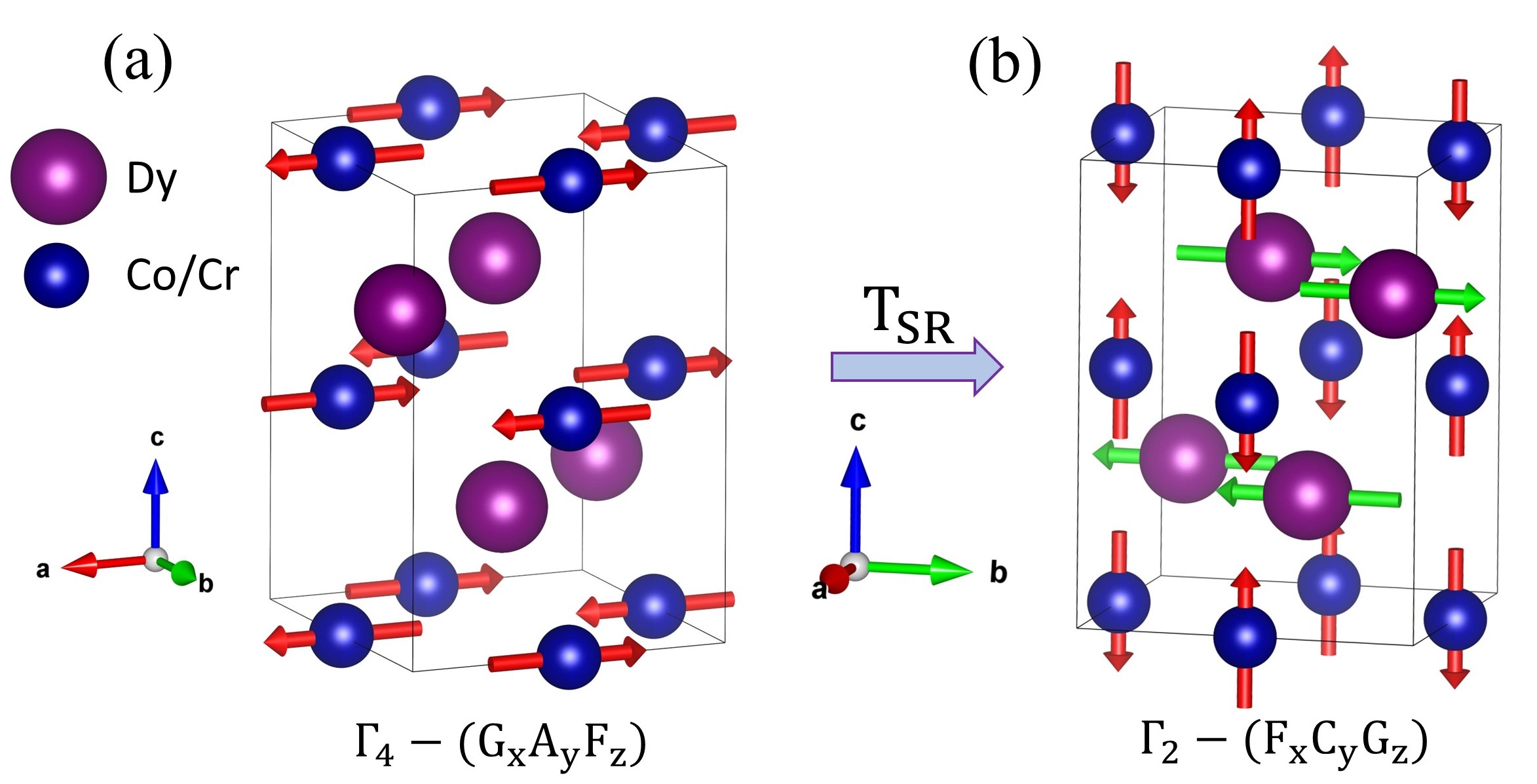}
 \caption{Transition of spin structure from (a) $\Gamma_4$(G$_x$A$_y$F$_z$) to (b) $\Gamma_2$(F$_x$C$_y$G$_z$)}\label{fig:spin}
\end{figure}

Figure ~\ref{fig:Magnetic}(a) shows the FC and ZFC magnetization, M(T), from 4 to 300 K in an applied magnetic field of 500 Oe. The ZFC curve almost traces the FC curve with small irreversibility $\bigtriangleup$M=M$_{FC}$-M$_{ZFC}$= 0.03 emu/g at 4 K attributed to the interfacial exchange interaction between feromagnetic (FM) and antiferomagnetic (AFM) states caused by the canted AFM ordering of DCCO \cite{selvadurai2015influence,coutinho2018influence}, the existance of FM and AFM states in DCCO will be discussed further in this section. The first derivative of the FC magnetization curve figure ~\ref{fig:Magnetic}(b) shows two transitions, occurring at 31 K and 119 K. The transition at temperature 119 K (T$_N$) is related to the canted AFM ordering of DCCO perovskite oxide owing to the anti-symmetric Dzyaloshinskii-Moriya interaction among Co and Cr ions, while Dy ions retrain their paramagnetic state \cite{sibanda2022structural,mazumdar2021structural}. With a decrease temperature below T$<$T$_N$, magnetization increases due to the continuous alignment of Dy ions in the internal field of $ \left|Co+Cr \right|$ ions before antiferromagnetic orders at spin reorientation temperature T$_{SR}$=31 K. Below T$_{SR}$=31 K, slope change in dM$_{FC}$/dT curve arise due to the reorientation of Dy$^{3+}$ moments antiparallel to the internal field of $ \left|Co+Cr \right|$ ions \cite{sibanda2022structural,mcdannald2013magnetic}. Most of the orthochromite except LaCrO$_3$ and most of the orthoferrite exhibit spin reorientation phenomena \cite{mazumdar2021structural,daniels2013structures}. The existence of spin reorientation phenomena can be explained by spin structure. There exist eight irreducible representations of space group, i.e., $\Gamma_1$- $\Gamma_8$; among them, only $\Gamma_1$(A$_x$G$_y$C$_z$), $\Gamma_2$(F$_x$C$_y$G$_z$), and $\Gamma_4$(G$_x$A$_y$F$_z$) are used to describe such spin structures \cite{mazumdar2021structural}. The net magnetic moment of $\Gamma_1$(A$_x$G$_y$C$_z$) is zero, but $\Gamma_2$ and $\Gamma_4$ show net FM moment along the x and z directions, respectively \cite{mazumdar2021structural}. Then spin reorientation takes place for the transition of spin structure from $\Gamma_4$(G$_x$A$_y$F$_z$) to $\Gamma_2$(F$_x$C$_y$G$_z$), as illustrated in figure ~\ref{fig:spin} \cite{sibanda2022structural,mazumdar2021structural,daniels2013structures}.

At temperatures above the transition temperature (T$_N$=119 K), the perovskite oxide material DCCO exhibits paramagnetic (PM) behavior, as illustrated in inset of figure ~\ref{fig:Magnetic}(b). To analyze the PM region of the inverse susceptibility curve, the Curie-Weiss (CW) law was employed for fitting using the equation $\chi$ = C/(T – $\theta$$_w$), where $\chi$, and $\theta$$_w$, and C are magnetic susceptibility, Weiss temperature and Curie constant, respectively \cite{das2017pr2fecro6,patra2022sign,hosen2023structural}. The parameters of slope and intercept obtained from the linear CW fitting correspond to the calculated values of C and $\theta$$_w$, respectively. From CW fitting, the Weiss temperature $\theta$$_w$ was found to be -22 K, which indicates the AFM ordering of DCCO perovskite oxide below T$_N$=119 K. Moreover, the experimental effective paramagnetic moment ($\mu _{eff}^{E}$) of DCCO perovskite oxide was calculated using the expression $ C=N_{A}\mu _{B}^{2}\mu _{eff}^{2}/3k_{B}$, resulting in a value of 15.21 $\mu$$_B$. This value is in good agreement with the theoretically calculated paramagnetic moment of 15.64 $\mu$$_B$ for the DCCO perovskite oxide, considering the individual paramagnetic moments of $\mu_{Dy^{3+}}$ = 10.6 $\mu_B$, $\mu_{Co^{3+}}$ = 4.9 $\mu_B$, and $\mu_{Cr^{3+}}$ = 3.88 $\mu$$_B$. In addition to the measurements taken with a 500 Oe applied field, we have also measured M-T curves at applied fields of 1000 Oe and 2000 Oe. These curves exhibit similar trends to the 500 Oe field but with higher magnetization values, as illustrated in the inset of figure ~\ref{fig:Magnetic}(a).

Furthermore, magnetic field-dependent magnetization (M-H) measurements were conducted using an applied magnetic field ranging from $\pm$ 7 T at temperatures of 20 K, 100 K, 200 K, and 300 K. The results of these measurements can be observed in figure ~\ref{fig:Magnetic}(c). The liner and unsaturated M-H curves obtained at 200, and 300 K demonstrate the PM nature of DCCO at these temperatures with no hysteresis loop \cite{bhuyan2021sol}. At temperature 100 K, the M-H curve was predominantly linear, though it exhibited a very narrow loop at its center. A magnified image of the M-H curve obtained at 100 K is displayed in the upper inset of figure ~\ref{fig:Magnetic}(c). From this M-H loop, the coercive field ($H_c$) and remanent magnetization ($M_r$) are found to be ~27 Oe and ~0.01 emu/g, respectively. This observation of the M-H curve at 100 K is also consistent with the M-T measurement. Notably,  at a lower temperature of 20 K, a narrow hysteresis loop was found, as depicted in the lower inset of figure ~\ref{fig:Magnetic}(c). It is worth noting that the loop did not reach full saturation even when subjected to a higher field of $\pm$ 7 T. The presence of this partially saturated hysteresis loop at lower temperatures suggests the coexistence of both FM and AFM states in the DCCO nanomaterial \cite{bhuyan2021sol,feng2014high}. The loop suggests FM interactions, whereas the partially saturated behavior up to 7 T indicates AFM interactions. The FM state in DCCO perovskite oxide originates from the 180$^{\circ}$ superexchange interaction between a half-filled Co$^{2+}$ d(e$_{g}^{2}$) orbital and an empty Cr$^{3+}$ d(e$_{g}^{0}$) orbital \cite{das2017pr2fecro6,hosen2023structural,kanamori1959superexchange}. Furthermore, this FM interaction transforms into AFM interaction due to lattice distortion existing in the DCCO \cite{feng2014high}. Particularly, the super-exchange interaction transforms from FM to AFM for a bond angle of 125$^{\circ}$ to 150$^{\circ}$ \cite{feng2014high}. In the case of DCCO, the average bond angle was determined to be 148$^{\circ}$ from XRD analysis, which is responsible for the AFM state along with the FM state.

\subsection{Optical properties}

\begin{figure*}
\centering
\includegraphics[width=150mm]{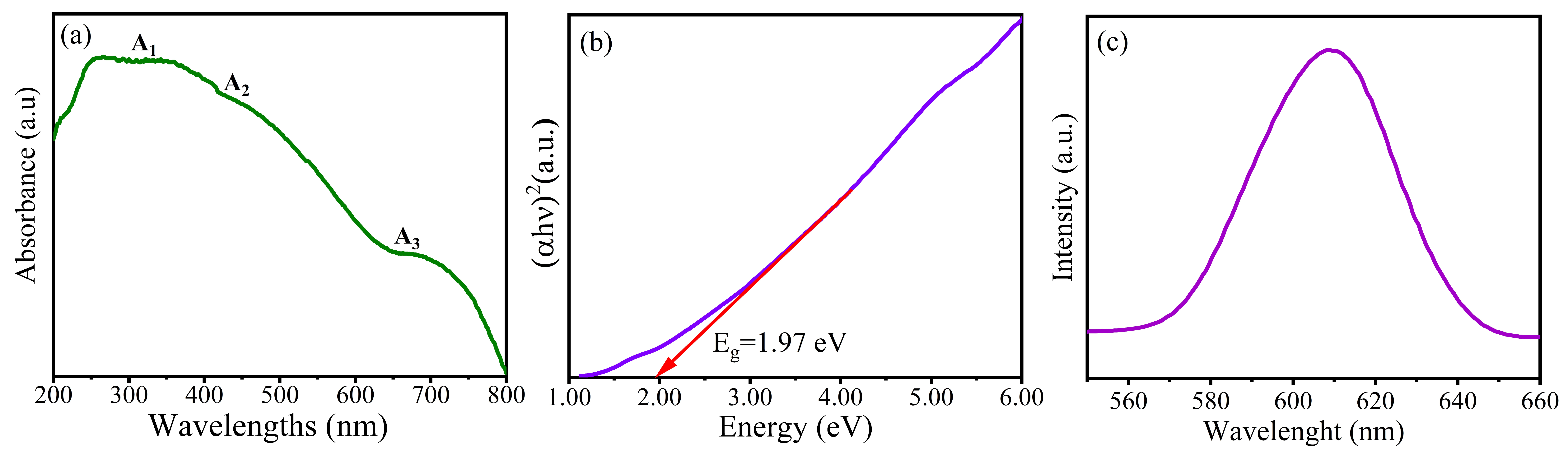}
\caption{(a) Spectrum illustrating UV-visible absorbance, (b) corresponding Tauc plot, and (c) steady state PL spectrum for Dy$_{2}$CoCrO$_6$}.\label{fig:UV}
\end{figure*}

The optical properties of as-prepared DCCO nanomaterials were studied extensively through UV–visible and photoluminescence spectroscopy. Figure ~\ref{fig:UV}(a) shows that the experimentally observed UV–visible absorbance spectrum revealed three absorption bands (labeled as A$_1$, A$_2$, and A$_3$) in the UV-visible range, confirming the multiband electronic structure of produced DCCO nanomaterials \cite{gaikwad2019structural}. The p-d charge transition [O(2p) → Co/Cr(3d)] in the Co/Cr-O octahedral center of DCCO nanoparticles is responsible for the absorption bands around 308 nm (A$_1$) and 433 nm (A$_2$) \cite{arima1993variation}. The low energy absorption band (A$_3$) near 640 nm is related to the p-p electronic transition in DCCO perovskite oxide \cite{gaikwad2019structural}. We next used the absorbance information to determine the optical bandgap of the DCCO perovskite oxide. Particularly, the bandgap of DCCO was calculated using the Tauc relation \cite{tauc1966optical,nowak2009determination}, which is given by
\begin{eqnarray}
 h\nu \alpha =A(h\nu -E_{g})^{n}
\end{eqnarray}
where, $h\nu$ is the incident photon energy, $\alpha$  is the coefficient of absorption, A is the constant of proportionality, n is the transition type, and E$_g$ is the bandgap energy. For the direct bandgap energy, the value of n is 1/2, and for the indirect bandgap n = 2, \cite{nowak2009determination}. As seen in figures ~\ref{fig:UV}(b), the direct bandgap of DCCO perovskite oxide found from extrapolation of the linear portion of the graph is 1.97 eV.

Moreover, we have performed steady-state PL spectroscopy to confirm the bandgap nature (either direct or indirect). Figure ~\ref{fig:UV}(c) shows the steady-state PL spectrum obtained from PL spectroscopy. As seen in figure ~\ref{fig:UV}(c), the PL peak of DCCO is at the wavelength of 611 nm, corresponding to a bandgap energy of 2.03 eV, which nearly matched the direct bandgap value of DCCO obtained from the UV-visible absorbance spectrum. For PL spectroscopy, the probability of radiative recombination decreases with time, and the probability of non-radiative recombination increases due to phonon emission \cite{bhuyan2021sol}. Moreover, steady-state PL spectroscopy ascribes only the radiative recombination of electron-hole pairs. Consequently, it is reasonable to assume that our as-prepared nanostructured DCCO is a direct bandgap material with a bandgap value of 1.97 eV. Significantly, this result illustrates the semiconducting behavior of DCCO perovskite oxide and, more crucially, exhibits its ability to absorb visible light from solar energy effectively.
Therefore, the potentiality of DCCO nanomaterials as a photocatalyst was examined by calculating the conduction-band-minima (CBM) and valence-band-maxima (VBM) potentials of DCCO nanomaterials through the Mulliken electronegativity calculation \cite{mulliken1934new,mulliken1935electronic}. According to Mulliken electronegativity approach,

\begin{figure}
\centering
\includegraphics[width=90 mm]{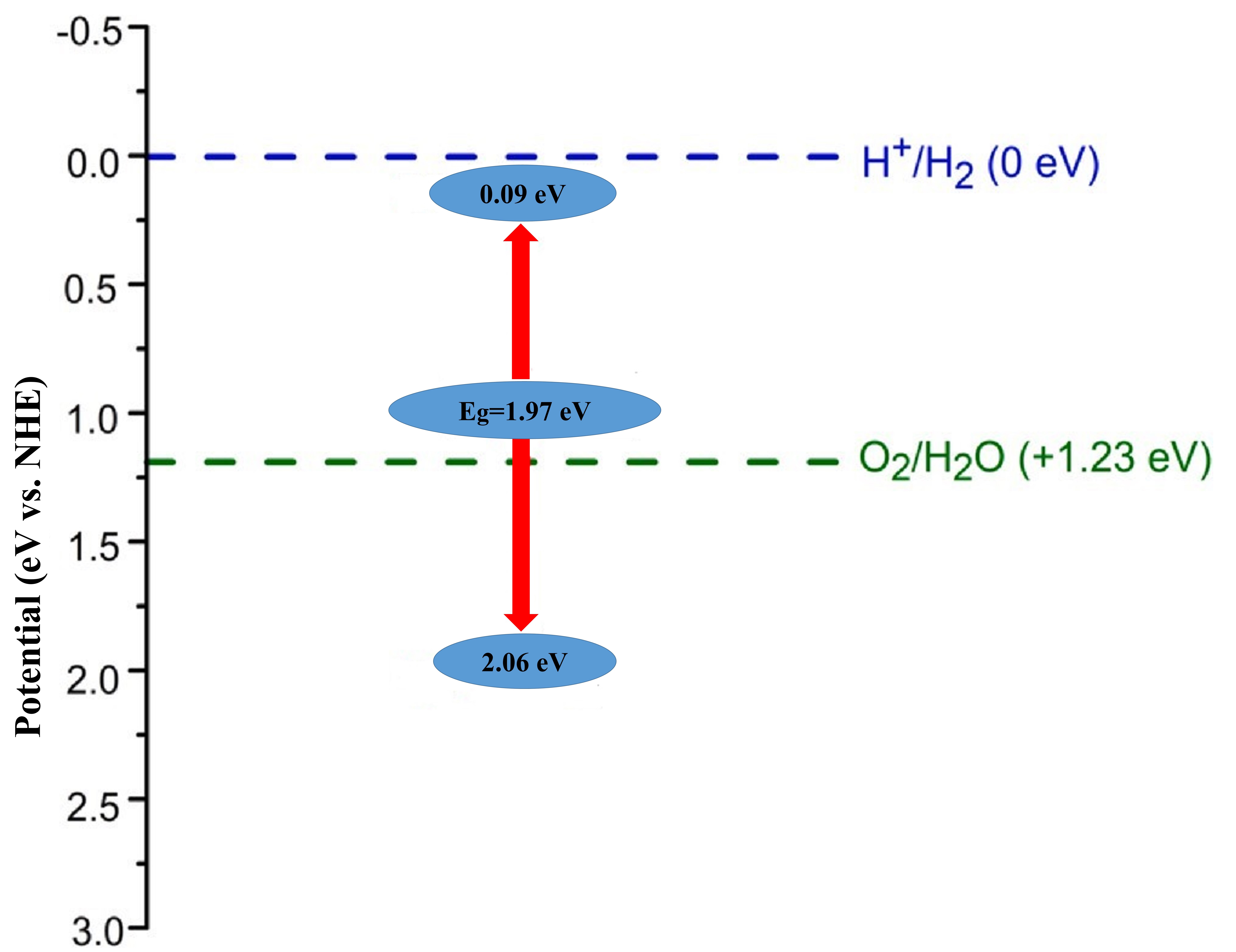}
\caption{Schematic representation illustrating the predicted energy band positions of nano-sized Dy$_{2}$CoCrO$_6$} particles.\label{fig:band position}
\end{figure}

\begin{eqnarray}
E_{CBM}=\chi-E_c-\frac{1}{2}E_g
\end{eqnarray}
\begin{eqnarray}
E_{VBM}=E_{CBM}+E_g
\end{eqnarray}
where $\chi$, E$_c$, and E$_g$ are Mulliken electronegativity (5.57 eV for DCCO perovskite oxide), free electrons energy (4.5 eV on normal hydrogen scale), and bandgap energy (1.97 eV obtained from Tauc plot). Then the values of CBM and VBM are found to be 0.09 eV and 2.06 eV, respectively. Figure ~\ref{fig:band position} shows the CBM and VBM band edge potential with respect to H$^+$/H$_2$ (0 eV) and O$_2$/H$_2$O (1.23 eV) potential on the standard hydrogen scale. Theoretically, a semiconductor material can produce photocatalytic hydrogen from water when CBM is more negative than H$^+$/H$_2$ potential and can produce oxygen from the water when VBM is more positive than O$_2$/H$_2$O potential \cite{hosen2023structural,lin2021photocatalytic}. As seen in figure ~\ref{fig:band position} the VBM potential of DCCO is more positive than O$_2$/H$_2$O potential, which reveals its promising potential for photocatalytic oxygen production from water \cite{bhuyan2021sol,lin2021photocatalytic}. Noticeably, it is difficult to produce metal-oxide semiconductors that can decompose water under visible light. Hence, DCCO nanomaterials can be a promising material for visible light-driven photocatalysts, particularly for photocatalytic O$_2$ evolution by splitting water due to its favorable bandgap of 1.97 eV and band edge potentials.

\section{Conclusions}
Structural, morphological, magnetic, and optical properties of sol-gel synthesized DCCO nanomaterials were investigated extensively. The absence of unwanted secondary peaks in the XRD spectrum showed the successful synthesis of DCCO nanoparticles. FTIR analysis confirmed the presence of NO$^{3-}$ ions, C-O, and O-H bonds in the DCCO nanomaterials. A morphological study also confirmed the successful synthesis of nanosized DCCO perovskite oxide particles with a mean particle size of 57 nm. Two transitions in dM$_{FC}$/dT curve demonstrated the Néel temperature, T$_N$ = 119 K, with spin reorientation temperature T$_{SR}$ = 31 K. The paramagnetic moment for DCCO perovskite oxide was found to be 15.21 $\mu$$_B$, which is consistent with the theoretical one. Hysteresis loop at low temperatures demonstrated the coexistence of FM and AFM states in the DCCO nanomaterials. Finally, the bandgap of DCCO was found to be 1.97 eV, which revealed the promising potential of DCCO nanomaterials in photocatalytic O$_2$ evolution from water. Magnetic and optical properties reveal that the DCCO nanoparticles are a strong candidate for next-generation spintronic devices and photocatalytic applications. 

\section*{Acknowledgments}
The financial assistance from the research cell, Mawlana Bhashani Science and Technology University is acknowledged. 

\section*{Data Availability}
The raw/processed data required to reproduce these findings cannot be shared at this time due to technical or time limitations.

\end{document}